\documentclass[aps,prl,showpacs,twocolumn]{revtex4}
\usepackage{graphicx}
\begin{document}
\newcommand {\be}{\begin{equation}}
\newcommand {\ee}{\end{equation}}
\newcommand {\bea}{\begin{eqnarray}}
\newcommand {\eea}{\end{eqnarray}}
\newcommand {\nn}{\nonumber}

\title{   A phason disordered two dimensional quantum antiferromagnet }

\author{ Attila Szallas }
\author{ Anuradha Jagannathan }
\affiliation{Laboratoire de Physique des Solides, CNRS-UMR 8502, Universit\'e
Paris-Sud, 91405 Orsay, France }

\author{ Stefan Wessel }
\affiliation{Institut f\"ur Theoretische Physik, Universit\"at Stuttgart, 70550 Stuttgart, Germany }

\date{\today}

\begin{abstract}
We examine a novel type of disorder in quantum antiferromagnets. Our
model consists of localized spins with antiferromagnetic exchanges
on a bipartite quasiperiodic structure, which is geometrically disordered in such
a way that no frustration is introduced.
In the limit of zero
disorder, the structure is the perfect Penrose
rhombus tiling. This tiling is progressively disordered by augmenting the
number of random ``phason flips" or local tile-reshuffling operations. The ground state remains N\'eel ordered, and we have studied its properties as a function of increasing
disorder using linear spin wave theory and quantum Monte Carlo. We
find that the ground state energy decreases,
indicating {\it enhanced} quantum fluctuations with increasing disorder. 
The magnon spectrum is progressively smoothed, and the
effective spin wave velocity of low energy magnons increases with
disorder. For large disorder, the ground state energy as well as the
average staggered magnetization tend towards limiting values
characteristic of this type of randomized tilings. 

\end{abstract}
\maketitle

This paper discusses the effects of a novel type of geometric disorder on the ground state of
Heisenberg antiferromagnets.  The clean system is  a perfectly deterministic quasiperiodic two-dimensional
tiling, and the type of disorder we consider is geometric, involving
a discrete shift of a randomly selected subset of sites. These
phason flips are operations that correspond to
reorganizing the structure locally, in the vicinity of the flipped
site. This type of disorder is strongly constrained, as one does not
modify the basic building blocks of the structure, but only the way
they are connected. For many quasicrystalline alloys,
samples of good quality are believed to possess long range order, with some thermally induced phonon
and phason disorder present at finite temperatures~\cite{stein}. However, one can also consider
essentially random quasiperiodic structures, which could be preferred for entropic reasons~\cite{henley}.

\begin{figure}[ht]
\begin{center}
\includegraphics[scale=0.5]{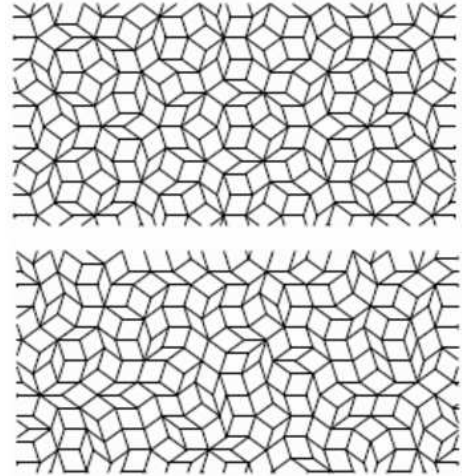}
\caption{Illustration of a perfect (upper panel) and a phason disordered Penrose tiling (lower panel).}
\label{tilings.fig}
\end{center}
\end{figure}
Quenched phason disorder has been considered previously in  other contexts, in particular concerning
electronic properties, and the possibility of Anderson localization in such systems was discussed.
Benza et al.~\cite{ben}
considered quantum diffusion in a two dimensional randomized tiling, and found that the typical value
of the diffusion exponent was larger for the disordered system compared to the pure case.
This means that quasicrystals show delocalization-like effects from disorder rather than the opposite,
weak localization due to disorder, observed in periodic systems.
Piechon and Jagannathan came to the same conclusion by analyzing the statistics of the
energy levels in phason disordered tilings (see e.g. Refs.~\cite{sire, piech_prb, piech_rev} for reviews).
Schwabe et al. \cite{schwa} analyzed the effect of phason flips on the electronic levels and wavefunctions, showing that they lead to a
smoothing of the density
of states, as well as of the fluctuations of the conductance. Finally, the effects of randomness on the
phonon spectrum has been considered in two dimensional quasiperiodic tilings (see e.g. Ref.~\cite{phonon}).

The study of the Heisenberg model on a quasiperiodic
tiling is motivated by  experimental findings of antiferromagnetic correlations in
quasiperiodic ZnMgR alloys (R: rare earth)~\cite{sato}. In  theoretical models
considered thus far~\cite{pl,anu, wess, prb1, prb2}  bipartite tilings are considered, with all sites
occupied by spins, and their exchanges restricted to adjacent
lattice sites, such that no frustration arises. Furthermore, these studies did not take into account the effects of disorder,
which is almost certainly also present in the experimentally studied alloys.
Here, we focus on the  effects that arise in quasiperiodic magnets in the presence of
phason disorder, which is
found to lead to enhanced quantum fluctuations, as indicated by the lowering of the ground state energy along with a reduction of the staggered magnetization with increasing disorder.

\begin{figure}[t]
\begin{center}
\includegraphics[scale=0.3]{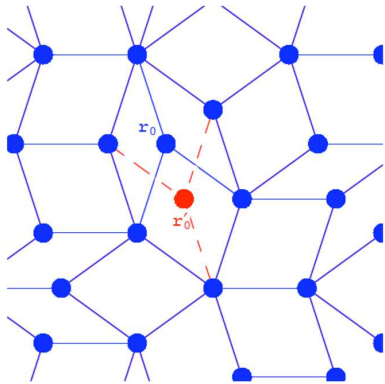}
\caption{(Color online)
A single phason flip, showing the original ($r_0$) and final ($r'_0$)
position of the shifted lattice site. New edges are shown by dashed
lines.}
\label{onephasonflip.fig}
\end{center}
\end{figure}
Fig.~\ref{tilings.fig} shows a portion of a perfect (deterministic) Penrose tiling and a typical example
of a phason disordered tiling. We consider in the following the
antiferromagnetic spin-1/2 Heisenberg model with Hamiltonian
$H=J \sum_{\langle i,j \rangle} {\mathbf S}_i \cdot {\mathbf S}_j$
for spins ${\mathbf S}_i$
located on all vertices of such tilings. Nearest-neighbor exchanges are antiferromagnetic $J>0$, and act between pairs of sites that are linked by an edge. Despite the fact that all couplings are equal, the ground state is spatially inhomogeneous contrarily to e.g. the square lattice antiferromagnet, where the staggered magnetization is uniform. The local staggered
magnetizations vary as a function of the local environment, and the ground state takes a complex,
hierarchically organized structure~\cite{pl, anu, wess, prb1, prb2}. The disorder considered here is purely geometric, i.e. the coupling  $J$ along bonds remains
fixed at a constant value.

Using linear spin wave theory (LSWT) and quantum Monte Carlo (QMC) calculations, we consider
periodic approximants of the Penrose tiling~\cite{prb1,prb2}. These are finite samples of $N$ spins,
satisfying periodic boundary conditions. Perfect tilings are obtained by the cut-and-project method (see e.g.~\cite{prb2})
after which they are disordered, by the
following method:
A phason flip is a process by which a 3-fold site hops to a
new allowed position (in terms of the tile configurations). The
old site disappears, as do the three bonds linking it to its
neighbors, while a new 3-fold site appears  on the
other sublattice (Fig.~\ref{onephasonflip.fig}). In our phason generating procedure, we randomly select a three-fold site.
If $\vec{r_0}$ and $\vec{r_i}$ ($j=1,2,3$) denote the position vectors
of the central site and its three neighbors, the new position of the site
is given by $\vec{r'_0}-\vec{r_0}=\sum_{j=1}^3
(\vec{r_j}-\vec{r_0})$ (Fig.~\ref{onephasonflip.fig}). Three new bonds appear linking
the new site to the sites at the positions $\vec{r'_j}$ ($j=1,2,3$).
The coordination numbers of all of the seven sites involved are
updated, and the whole procedure is repeated, with the constraint that there be an equal number of flipped sites on the $A$ and the $B$
sublattice, so as to preserve the condition $N_A=N_B$, where $N_A$ ($N_B$) denotes the number of sites of the $A$ ($B$) sublattice.

\begin{figure}[t]
\begin{center}
\includegraphics[scale=0.65]{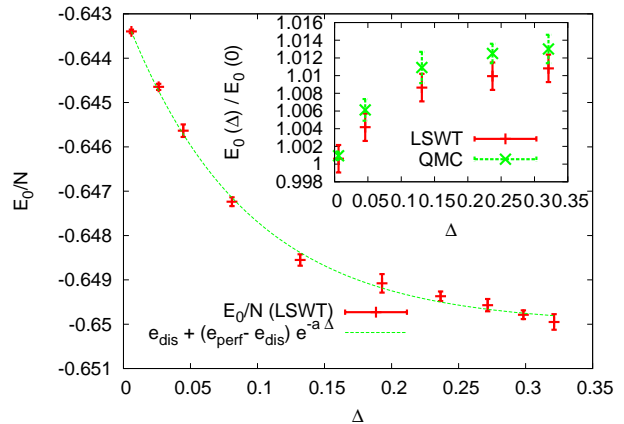}
\caption{ (Color online) Dependence of the ground state energy on the disorder strength $\Delta$ calculated within LSWT.
The smooth curve is a fit to an exponential decay towards a higher disorder value. The
system size $N$=4414. The inset shows the normalized ground state energy as a function of disorder $\Delta$. Points represent values obtained after
scaling LSWT and QMC data to infinite system sizes.}
\label{GSE.fig}
\end{center}
\end{figure}

For a given total number of phason flips $N_{ph}$, the degree of disorder $\Delta$ is defined as the average overlap distance between the perfect sample and the
disordered samples. The overlap for a given sample is defined by $N_{ph}^{-1} \sum_{i=1}^{N_{ph}} \eta_i$, where $\eta_i$ is 1 or 0, depending on whether  site $i$ is shifted or not  with respect to the perfect tiling. The whole procedure is then repeated for
approximants of system sizes $N=246$, $644$, $1686$ and $4414$ in LSWT and $246$, $644$ and $1686$ in QMC. Averaging is carried out over
a large number of samples: 100 samples for $N=246$, $644$, $1686$ and $10$ samples for $N=4414$ in LSWT and $40$ samples for $N=246$ and $20$
samples
for
$N=644$, $1686$ in the
QMC.
This
method of disordering does not change the overall number of rhombi of each kind, and generates samples of fixed phason strain~\cite{note}. Due to the fact that we obtain  a defected structure (a periodic approximant, which is only locally equivalent to the infinite quasiperiodic tiling) there is an upper limit to the number of phason flips we are able to introduce in the tilings. Unacceptable configurations are  observed to appear after a number of flips larger than about
$1.3 N$, corresponding to the number of steps after which the periodic boundary conditions are felt by the system.

After the randomized samples are obtained, LSWT
calculations are performed as described in~\cite{wess, prb2}.
Spin operators are transformed using the Holstein-Primakoff
transformation to bosonic operators $a_i$, $b_j$ $(i,j=1,...,N/2)$, corresponding the $A$ and $B$ sublattices respectively.
The linearized Hamiltonian in the boson operators is then
diagonalized numerically. Once the eigenmodes have been
determined, one obtains the ground state energy, and local
staggered magnetizations for each realization of disorder. Finally, we carry
out the statistical analysis of the results, by performing disorder averaging.
Within the QMC simulations, we obtain the local staggered magnetization
$m^2_{s}(i)=\frac{3}{N}\sum_{j=1}^{N} \epsilon_i \epsilon_j \langle S^z_i
S^z_j \rangle$ from the spin-spin correlation
function~\cite{pl, wess}, where $\epsilon_i\pm 1$, depending on whether lattice site $i$ belongs to sublattice $A$ or $B$. The QMC
simulations
were
performed using
the
stochastic series expansion method~\cite{sandvik}
at temperatures taken low enough to obtain ground
state properties of these finite systems~\cite{wess}.

\begin{figure}[t]
\begin{center}
\includegraphics[scale=0.65]{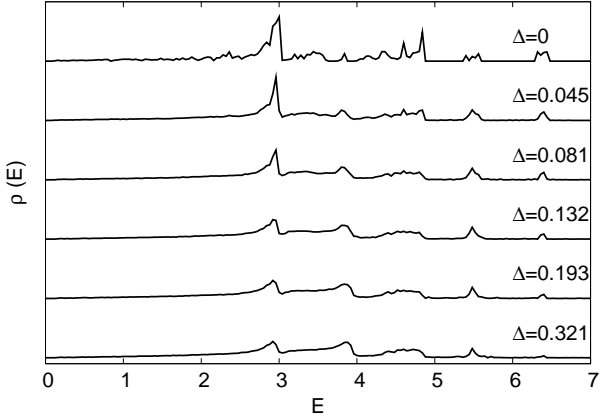}
\caption{ Evolution of the density of states $\rho(E)$ for various values
of the disorder strength $\Delta$ from LSWT on the $N=4414$ sites system.}
\label{dos.fig}
\end{center}
\end{figure}

We now describe our results on the effect of phason disorder.
First, we consider the
average ground state energy per site $E_0/N$, shown as a function of disorder in
Fig.~\ref{GSE.fig} for the $N = 4414$ sites system. The values have been normalized with respect to the value
obtained for the clean system. Also shown is  a fit to an exponential decay towards a limiting value in the
strong disorder case, of
the form $E_0/N=e_{dis} + (e_{perf}- e_{dis}) e^{-a\Delta}$, with $a=11.16$, $e_{dis}=-0.6500$ and
$e_{perf}=-0.6429$.
The asymptotic value of the ground state energy $e_{dis}$ represents the average value of the ground state energy of maximally randomized Penrose approximants,
which lies
below the ground state energy of the perfect system.
This indicates that the introduction of phasons tends to enhance quantum fluctuations in the tilings, as
compared to the clean case.
The inset of Fig.~\ref{GSE.fig} shows the ground state energy per site as a function of disorder, obtained from  LSWT and QMC simulations,
normalized with respect to the value obtained for the clean system.
The results obtained from LSWT and QMC are found to be in good agreement. This also shows the applicability of
the linear spin wave approximations to this randomized system.

\begin{figure}[t]
\begin{center}
\includegraphics[scale=0.32]{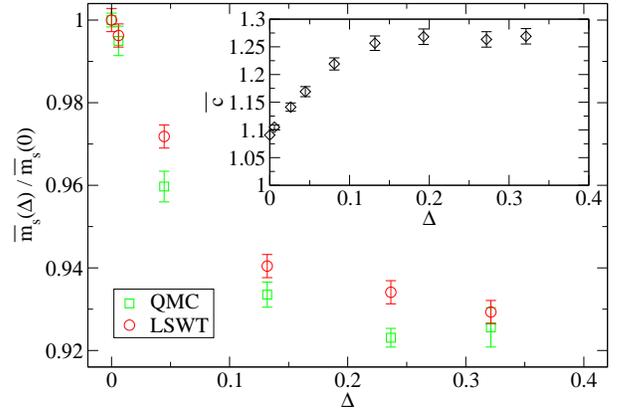}
\caption{(Color online) Normalized averaged staggered magnetization as a function of phason disorder strength $\Delta$.
Points represent values obtained from LSWT and QMC data after scaling to the thermodynamic limit.
The inset shows the averaged spin wave velocity $\bar{c}$ as a function of
the disorder strength $\Delta$ as obtained within LSWT.}
\label{avgmag.fig}
\end{center}
\end{figure}

Next, we consider
the density of states, defined via
$\rho(E) = \sum_\mu \delta(E - \omega_\mu)$, and shown in Fig.~\ref{dos.fig}.
We find that the dominant effect of phason disorder on the density of states is to progressively
smoothen fluctuations and
fill in gaps, as  in the electronic case~\cite{sire}. The low energy tail is quadratic, and can be fitted
to obtain the averaged low-energy spin wave velocity $\bar{c}$, that {\it increases} with $\Delta$ as shown in the inset of
Fig.~\ref{avgmag.fig}.
This indicates that spin wave
propagation is facilitated by the
phason disorder, in analogy with the problem of quantum diffusion of
electrons in the tight binding model in quasiperiodic tilings~\cite{sire}.
In addition, the localized states at $E=3$ disappear
progressively. These states arise on closed
loops of 3-fold sites~\cite{prb2}. They are hence destroyed when a phason flip
occurs on one of the participating sites.

\begin{figure}[h]
\begin{center}
\includegraphics[scale=0.7]{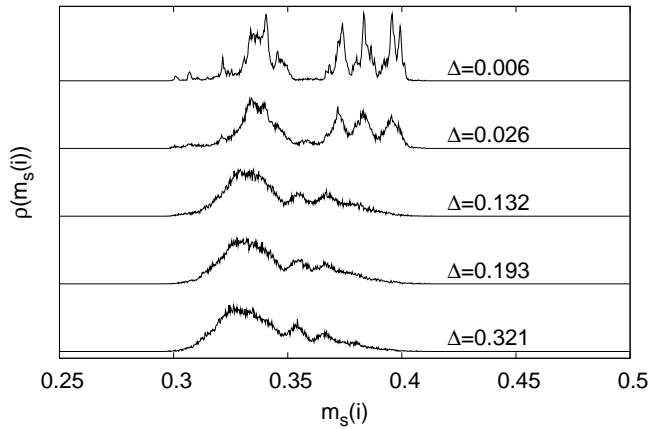}
\vspace{.2cm} \caption{Evolution of the distribution in the local staggered magnetization with the disorder strength, as obtained within LSWT
on the 4414 sites system.}
\label{magdist.fig}
\end{center}
\end{figure}

Finally, we turn to discuss the evolution of the staggered magnetization upon introducing phason disorder.
Fig.~\ref{avgmag.fig} shows the spatial average of the staggered
magnetization (i.e. after averaging over all of the sites) as a function of disorder, normalized with respect to the value obtained for the clean system. The curve
shows a clear  decrease in the global staggered magnetization with
increasing disorder. As for the ground state energy curve, the decrease eventually levels off.
It is also interesting to analyze the evolution of the full distribution of the staggered magnetizations with increasing phason disorder.
In the perfect tiling, this distribution has several peaks, each
of which corresponds to a distinct coordination number. In the disordered tilings, the differences between the coordination number is smoothed out. As Fig.~\ref{magdist.fig} shows, the distribution becomes smoother, as the  disorder is increased. In addition, the average value shifts to lower values. The smoothing occurs due to a larger number of local environments created by the phason
flips, and due to the loss of self similarity on larger
length scales.

\begin{figure}[t]
\begin{center}
\includegraphics[scale=0.9]{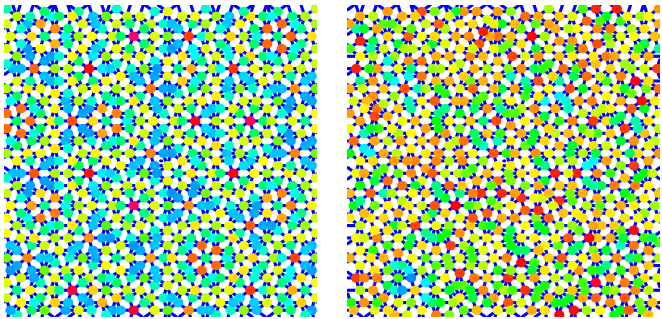}
\vspace{.2cm} \caption{(Color online) Color representation of the local staggered magnetizations in a perfect tiling (left) and a disordered tiling (right) showing sites of magnetizations ranging from high (blue) to low (red) values.} \label{compare.fig}
\end{center}
\end{figure}

Fig.~\ref{compare.fig}
shows,  that the distribution is more homogeneous in the disordered
tiling, which furthermore lacks the hierarchical features of the
perfect tiling. The redder appearance of the right  figure also
illustrates the already noted fact, that globally the staggered magnetization
is lower in the disordered system.

In conclusion, we analyzed the effects of a new type of geometrical
disorder on the magnetic properties of  quasiperiodic
antiferromagnets. Concerning the low energy modes, magnons propagate
with a slightly higher velocity in the disordered tiling as compared
to the perfect quasiperiodic tiling. The density of states is
smoothed. Degenerate states localized on closed loops disappear with
increasing disorder. Eigenmodes tend to become more delocalized as
compared to the perfect tiling. These effects have their analogy in
phonon models, as well as in the tight binding model for electrons
in quasiperiodic tilings. In the magnetic problem our results also
indicate that the effect of disorder is to reduce the strong
coherent backscattering of the magnon wavefunctions due to a perfect
quasiperiodic potential and to favor a more diffusive dynamics. Upon
increasing phason disorder in the antiferromagnetic model, the
ground state progressively loses its self-similar features, and
tends towards a more homogeneous distribution of staggered moments.
The global average of the staggered magnetization decreases, as does
the ground state energy, signaling increased quantum spin
fluctuations. Both quantities tend towards a limiting value for this
class of disordered tilings. Such two
dimensional quantum antiferromagnets may be experimentally realizable in the near future,
as there has been considerable progress recently with depositing
atoms on quasiperiodic surfaces, which serve as templates. A monolayer of
Pb atoms having quasiperiodic symmetry has 
been obtained \cite{ledieu}. To study quantum magnetism one
needs to obtain such a surface layer using low spin atoms. In addition, one needs the interactions to be predominantly short range
and antiferromagnetic, and this could be realized via a
superexchange mechanism, as in the cuprate
layers of high $T_c$ compounds, in which oxygen atoms mediate the exchange between the $S=1/2$ copper spins.

\acknowledgments We thank HLRS Stuttgart and NIC J\"ulich for allocation of computing time. A.S. was supported by the European Commission, through a Marie Curie Foundation contract, MEST CT 2004-51-4307, during the course of this work.


\begin{references}
\bibitem{note} In the cut-and-project scheme (see e.g.~\cite{prb2}) the deviations from perfect order correspond to fluctuations in which the orientation of the
cut in five dimensional space remains constant.
\end{references}
\end{document}